\documentclass[twocolumn,           
               showpacs,            
               nopreprintnumbers,     
               aps,                 
               prl,          	    
               letterpaper,             
               groupeaddress,      
               nofootinbib,         
               tightenlines,        
               floats,floatfix      
               ]{revtex4-1}
\usepackage{graphicx}
\usepackage{dcolumn}
\usepackage{bm}
\usepackage{mciteplus}
\usepackage{amsmath}
\usepackage{amsfonts,amssymb}
\usepackage{color}
\usepackage{enumerate}

\begin{document}

\title{An Effective Cosmological Constant from an Entropic Formulation of Gravity}
\author{I. D\'iaz-Salda\~na$^{1}$}
\author{J. L\'opez-Dom\'inguez$^{2}$}
\author{M. Sabido$^{1,3}$}
\affiliation{$^{1}$Departamento de F\'isica, Divisi\'on de Ciencias e Ingenier\'ias, Universidad de  Guanajuato, Campus Le\'on, Loma del Bosque 103, Lomas del Campestre, C.P. 37150, Le\'on, Gto.}
\affiliation{$^{2}$Unidad Acad\'emica de F\'isica, Universidad Aut\'onoma de Zacatecas, Calzada Solidaridad esquina con Paseo a la Bufa S/N C.P. 98060, Zacatecas, M\'exico.}
\affiliation{$^3$ Department of Theoretical Physics, University of the Basque Country UPV/EHU, P.O BOX 644, 48080 Bilbao, Spain.}

\begin{abstract}

We use the ideas of entropic gravity to derive  the FRW cosmological model and show that for late time evolution we have an effective cosmological constant. By using the first law of thermodynamics and  the modified entropy area relationship derived from the supersymmetric Wheeler-DeWitt equation of the Schwarzschild black hole, we obtain modifications to the Friedmann equations that in the late time regime gives an effective positive cosmological constant. Therefore, this simple model can account for  the dark energy component of the universe by providing an entropic origin to the cosmological constant $\Lambda$.

\end{abstract}

\draft
\date{\today}
\maketitle

\section{Introduction} \label{Int}

 There is a puzzling observations that has the potential of fundamentally altering the foundations of physics. This observation came at the end of the millennium, and revealed that the Universe is in a period of accelerated expansion. Explaining the origin of this acceleration has encouraged the formulation of  hypothesis that clash with the mainstream theories of physics. The several proposals that exist to solve what is currently known as the dark energy  problem, can be divided in two clases. In the first class we remain in the context of general relativity (GR), an propose that this acceleration is produced by a unknown energy density or to the existence of a new type of matter that has the nonphysical property of negative pressure. In this context, it seems that the best candidate is the cosmological constant $\Lambda$. Even if the observations are compatible with the existence of $\Lambda$ there are serious theoretical problems to consider this model as the final answer \cite{weinberg,lambda}. The second class, assumes  that we do not need a new energy density with strange physical properties, and considers that this  acceleration is a consequence of a modified theory of gravity. For the last decade, there has been a lot of work in modified theories of gravity \cite{padilla1}, constructing  models that exhibit  an accelerating scale factor in their late time dynamics \cite{padilla2}. The most successful attempts are Horndenski  type theories \cite{Horndeski}, these are scalar tensor theories that have second order field equations and are free of ghost type instabilities \cite{18}. 
In summary one can address the dark energy problem by either, staying in the framework of GR and changing the nature of the energy density by introducing a dark energy component, or be confident on our knowledge of the energy and matter content of the universe and modify the gravitational sector, and therefore use an alternative to GR.

 If we entertain the possibility that the dark energy problem is  a consequence of the poor understanding of gravity, we need to formulate an alternative to GR.
 A recent approach to understand the incompatibility of gravity with quantum mechanics is to consider gravity  not as fundamental interaction but as an emergent phenomenon \cite{Jacobson:1995ab}. The renewed interest in this idea started with Verlinde \cite{Verlinde:2010hp}, where the author claims that Newtonian gravity is an entropic force, in analogy with the emergent forces that are present in the study of polymers. Since in this formulation Newtonian gravity has an entropic origin, one can propose modifications to Newtonian gravity by analyzing modifications to the Bekenstein-Hawking entropy-area relation. 
 In a more recent paper \cite{Verlinde2}, Verlinde explores the possibility of a common origin of dark matter  and dark energy in the context of an emergent formulation of gravity. 
The dark matter predictions for this theory have been put to test in several works \cite{niz,test2}. However, even if the specific formulation of Verlinde is not the last answer, the entropic or emergent origin of gravity and its connection to the dark matter an dark energy can shed some light to origin of the dark sector of the Universe.

In this work we present a simple model that attempts to understand the  origin of dark energy from an alternative formulation of gravity. This approach can be understood in the context of an entropic formulation of gravity. We use the modified entropy-area relation that was obtained by applying the Feynman-Hibbs path integral procedure \cite{feynman}, to calculate  the entropy from the supersymmetric generalization of the Wheeler-DeWitt (WDW) equation for Schwarzschild black hole \cite{isaac}. Using the modified area-entropy relation one can obtain a modified Friedmann equation. The modifications to the Friedmann equation can be traced to the new terms that appear in entropy-area relation. From the modified Friedmann equation in  the late-time limit  we find  a de Sitter scale factor and therefore  able to define an effective cosmological constant. Finally, we can calculate the value of the free parameter of our model in order to be consistent with the current value of cosmological constant.  
 

\section{A New entropy-area relationship} \label{AE relationship}
For this alternative formulation of gravity, the main ingredient is the modified entropy-area relationship.
We begin by discussing  the entropy-area relationship obtained in \cite{isaac},   that we will be using to derive the modified Friedmann equations. 

This relation is obtained from the supersymmetric generalization of the WDW equation for the Schwarzschild black hole. Starting from the
Schwarzschild black hole metric,
for $r<2m$, $g_{rr}$ and $g_{tt}$, the components of the metric change  sign and $\partial_{t}$ becomes a space-like vector. {Hence} if we perform the transformation $t\leftrightarrow r$, 
and compare with the Misner parametrization of the Kantowski-Sachs (KS) metric
we can establish the diffeomorphism with KS.  Using this relationship between the Schwarzschild and KS   in the Einstein-Hilbert action and performing an integration over the spatial coordinates, we get an effective Lagrangian. Now is straight forward to obtain the Hamiltonian constraint and after canonical quantization  derive the WDW equation. 
The supersymmetric generalization of the supersymmetric quantum Schwarzschild black hole is 
\begin{eqnarray}\label{susyec}
&&\left[ -\frac{\partial^2}{\partial\Omega^2}+\frac{\partial^2}{\partial\xi^2}\right. \\
&&\left. + 12\left( 4\pm\frac{1}{\sqrt{e^{-2\sqrt{3}\Omega}+\epsilon^{-2/3}}} \right)e^{-2\sqrt{3}\Omega} \right] \Psi_{\pm}=0.\nonumber
\end{eqnarray}
The wave function has four components, although only two are linearly independent. The contributions of supersymmetry are encoded in the modified potential, where the quantity $\epsilon$ is the free parameter of our model, and  in the limit $\epsilon\to 0$ one recovers the original WDW equation. 
The modified entropy-area relationship is obtained by using the Feynman-Hibbs procedure~\cite{feynman}.  This approach is based on exploiting the similarities of the expression of the density matrix and the kernel of Feynman's path integral approach to quantum mechanics. By performing a Wick rotation $t\to i\beta$, we get the Boltzmann factor, and the  kernel is transformed to the density matrix. The potential to a first-order approximation can be written as $V(x)\approx V(x_1)$ for all the contributing paths.
 In this approximation, we can formally establish a map from the path integral formulation of quantum mechanics to the classical canonical partition function. The partition function is calculated in a classical manner, but using the corrected potential, and the quantum effects are encoded in the corrected potential which is a mean value of the potential $V(x)$ averaged over points near $\tilde x$ with a Gaussian distribution. 
According to this procedure, the partition function for the model in Eq.\eqref{susyec} is
\begin{eqnarray}
\mathcal{Z}(\beta)&=&\sqrt{\frac{2 \pi}{3}}\frac{1}{\beta E_{P}}\exp{\left ( -\frac{\beta^{2}E_{P}^{2}}{16\pi}-\frac{\beta^{3}E_{P}^{3}\epsilon}{96} \right )}\nonumber\\
&\times& \left ( 1+\frac{\pi \beta E_{P} \epsilon}{3} \right )^{-1/2}, \label{FPSUSY}
\end{eqnarray}
from which the entropy is calculated straightforwardly as $S/k_B=\ln{\mathcal{Z}}+\beta \bar{E}$, resulting in the following entropy-area relation 
\begin{eqnarray}
\frac{S(A)}{k_B}&=&\Delta(\epsilon)\frac{A}{4l_{P}^{2}}-\frac{1}{2}\ln{\frac{A}{4l_{P}^{2}}}\\
&+&\Gamma(\epsilon)\left(\frac{A}{4l_{P}^{2}}\right)^{1/2}+\epsilon\left(\frac{A}{2l_{P}^{2}}\right)^{3/2},\nonumber\label{entr}
\end{eqnarray}
where we defined the functions  $\Delta(\epsilon)= 1-10\epsilon^2+4\epsilon^4$ and  $\Gamma(\epsilon)=\sqrt{2}/3(-12\epsilon+23\epsilon^3-20 \epsilon^5+4 \epsilon^7)$.

{In this modified entropy-area relation we get several terms, among them a term  linear on the area $A$, that is the usual term for gravity. We also identify a term proportional to $A^{3/2}$, as we will see it will be the most relevant term for our work. Entropy terms that scale on the volume are typically related to ordinary degrees of freedom of quantum field theories and it is not common in gravity theories. It is worth mentioning that volumetric corrections  to the entropy of black holes has been derived  in loop quantum gravity \cite{Livine}. With respect to the remaining terms, the one proportional to $A^{1/2}$ exhibits the behavior of a self-gravitating gas, which is a system  with an entropy that scales as the characteristic length of the system \cite{deVega:2001zk}. Finally, the logarithmic correction  to the entropy has been derived from different approaches of quantum gravity applied to black holes \cite{Obregon:2000zd,Domagala:2004jt,Mukherji:2002de,Sen:2012dw}.}
\section{The modified Friedmann equation}\label{eg}
In previous works a deep relationship between the entropic forces and Newtonian gravity was conjectured. It was proposed that gravity is an effective force that has an entropic origin. This has given rise to several models that attempt to modify Newtonian gravity by introducing quantum modification to the entropy  \cite{Modesto:2010rm} or using alternative definitions of entropy \cite{Martinez-Merino:2017xzn}, in order to find modifications to Newton's gravitational force and also modifications to gravity in the cosmological scenario \cite{Sheykhi:2010yq}.
The insight to derive Einstein's equations from the relationship between entropy and gravity  was first proposed by Jacobson \cite{Jacobson:1995ab}, and more recently Verlinde used this idea to derive Newton's gravitational force. By relating the entropy with the information contained in a surface $\mathcal{S}$ that surrounds a mass $M$ and that is very near to a test mass $m$, an entropic derivation of Newtonian gravity can be obtained \cite{Verlinde:2010hp}. Therefore, we can find modifications to Newtonian gravity by analyzing modifications to the entropy-area relation. Following this formulation of gravity, in \cite{isaac}, the modified Newtonian force associated to Eq.(\ref{entr}) turns to be
 \begin{eqnarray}
\mathbf{F}_{M}&=&-\frac{G_{eff}Mm}{R^{2}}\left[1+\frac{3\sqrt{2\pi}  }{\Delta(\epsilon)l_{P}} \epsilon R \right.\\
&+&\left.\frac{l_{P}\Gamma(\epsilon)}{2\sqrt{\pi}\Delta(\epsilon)}\frac{1}{R}
-\frac{l_{P}^{2}}{2\pi\Delta(\epsilon)}\frac{1}{R^2}\right]\mathbf{\hat{R}},\nonumber \label{modforce}
\end{eqnarray}
where $G_{eff}=\Delta(\epsilon)G$ is the {effective} gravitational constant. 
This force has some interesting phenomenological consequences,
in particular, if we consider a star in  circular motion around the center of a galaxy, the velocity in the limit of large radius results in a constant velocity given by
$v^2\approx\frac{3\sqrt{2\pi}G M}{l_P}\epsilon$.   
Therefore we can argue that the entropic force contains the usual Newtonian gravitational force as well as a {\it ``dark force"} component that could 
account for the anomalous galactic rotation curves in the context of entropic formulation of gravity. 

Now we ask ourselves, can the entropy-area relationship in Eq.(\ref{entr}) in conjunction with the idea of entropic gravity, shed some light on the origin of dark energy? In order to answer this question we must analyze the entropy-area relationship, Eq.(\ref{entr}), in the cosmological scenario. 
It is known that imposing the cosmological principle is the basic requirement to obtain the FRW cosmological model, the resulting equation is the well known Friedmann equation which describes the dynamics of the Universe. {Introducing different components to the matter sector in the Friedmann equation allows to explain several aspects of the Universe \cite{Gorkavyi,Pardede,MaKai}. On the other hand we can use a different approach, obtain a modified Friedmann equation from the entropy-area relationship to study the dynamics of the Universe. Fortunately in this scenario the features of emergent gravity provide a natural and direct manner to obtain modifications to the Friedmann equation.} Starting from the entropy-area relation  which (at least in this context) is where gravity emerges, the modifications to Friedmann equation arise in a straightforward manner. In particular, we will follow the approach in \cite{Cai:2005ra}, where the Friedmann equation is obtained by using the first law of thermodynamics  on the so called apparent horizon of a FRW cosmological model. 
This approach was presented in \cite{cai}, where the authors consider several models. They first use an entropy with the typical logarithmic correction, and then a more general entropy that includes  a correction term that is proportional to $1/A$. Finally the method is generalized for the case of an entropy with a modification being an arbitrary function of the area.  

Let us proceed with the derivation of the Friedmann equation. As we now, the FRW universe is described by the metric
\begin{equation}
ds^{2}=-dt^{2}+a^{2}(t)\left( \frac{dr^{2}}{1-kr^{2}}+r^{2}d{\Omega}^{2}\right),
\end{equation}
comparing with
$
ds^{2}=h_{ab}dx^{a}dx^{b} +\tilde{r}^{2}d\Omega^{2},
$
we can identify $h_{ab}$.
Following the procedure in \cite{Cai:2005ra}, we introduce the work density $W$ and the  energy-supply vector $\Psi$
\begin{equation}
W=-\frac{1}{2}T^{ab}h_{ab},\quad \Psi_{a}=T^{b}_{a}\partial_{b}\tilde{r}+W\partial_{a}\tilde{r},
\end{equation}
where $\tilde{r}=a(t)r$ and $T_{ab}$ is the projection of $T_{\mu\nu}$ in the normal direction of the 2-sphere.  
 For a perfect fluid we get
\begin{eqnarray}
W&=&\frac{1}{2}(\rho-P),\\
\Psi_{a}&=&\left( -\frac{1}{2}(\rho+P)H\tilde{r},\frac{1}{2}(\rho+P)a\right).\nonumber
\end{eqnarray}
The amount of energy $\delta Q$ crossing the apparent horizon during the time interval $dt$ is given  by
\begin{equation}
\delta{Q}=-A\Psi=A(\rho+P)H\tilde{r}_{A}dt,
\end{equation}
where $\tilde{r}_{A}$ is the radius of the apparent horizon, $A$ is the area and is  given by $A=4\pi\tilde{r}^2_A$. From $h^{ab}\partial_{a}\tilde{r}\partial_{b}\tilde{r}=0$ we obtain
\begin{equation}
\tilde{r}^2_{A}=\left(H^2+\kappa/a^2\right)^{-1}.
\end{equation}
\noindent Using the Clausius relation 
$\delta Q= TdS,$   
the continuity equation  $\dot{\rho}+3 H(\rho+P)=0$, the modified entropy-area relationship of Eq.(\ref{entr}) 
and the time derivative of $A$, we get
\begin{eqnarray}
\frac{4\pi G}{3}\dot{\rho}&=&{\Delta}(\epsilon)\left (1-\frac{2G}{{\Delta}(\epsilon)}\frac{1}{A}+\frac{\sqrt{G}}{{\Delta}(\epsilon)}\Gamma(\epsilon)\frac{1}{A^{1/2}}\right. \nonumber\\
&+&\left. \frac{3\epsilon}{\sqrt{2G}{\Delta}(\epsilon)}  A^{1/2} \right)  H \left(\dot{H}-\frac{\kappa}{a^2} \right). \label{nonintr}
\end{eqnarray}
To obtain the Friedmann equation, we simply integrate Eq.(\ref{nonintr}) and use $2 H(\dot{H}-{\kappa}/{a^2})dt=d[H^2+\kappa/a^2]$
to finally at
\begin{eqnarray}\label{mfe}
&&\frac{8\pi G}{3}\rho={\Delta}(\epsilon)\left( H^2+\frac{\kappa}{a^2}\right)-\frac{G}{4\pi}\left( H^2+\frac{\kappa}{a^2}\right)^{2}\\ \nonumber
 &+&\frac{1}{3}\sqrt{\frac{G}{\pi}}\Gamma(\epsilon)\left( H^2+\frac{\kappa}{a^2}\right)^{3/2}+6\sqrt{\frac{{2\pi}}{G}}\epsilon\left( H^2+\frac{\kappa}{a^2}\right)^{1/2},
\end{eqnarray}
this is the modified Friedmann equation.
To study the dynamics of the Universe we simply need to introduce the appropriate matter, (i.e, radiation, dust, scalar field, etc.) and  solve the resulting equations. 
It is important to recall that in the units we are employing we have $G=l_{P}^{2}$. Considering this, let us note that the second and third terms in the r.h.s. of Eq.(\ref{mfe}) can be neglected, resulting in the modified Friedmann equation 
\begin{equation}
\frac{8\pi G}{3}\rho={\Delta}(\epsilon)\left( H^2+\frac{\kappa}{a^2}\right)+6\sqrt{\frac{{2\pi}}{G}}\epsilon\left( H^2+\frac{\kappa}{a^2}\right)^{1/2}. \label{modfrieddom}
\end{equation}
we can solve the last equation for $ H^2+\kappa /a^2$
\begin{eqnarray}\label{sol}
&&H^2+\frac{\kappa}{a^2}=\frac{8\pi G}{3}\frac{\rho}{{\Delta}(\epsilon)}+\frac{1}{2}\left(6\sqrt{\frac{{2\pi}}{G}}\frac{\epsilon}{{\Delta}(\epsilon)} \right) ^2\\
&&\pm \sqrt{\left(6\sqrt{\frac{{2\pi}}{G}}\frac{\epsilon}{{\Delta}(\epsilon)}\right)^4+4\left(6\sqrt{\frac{{2\pi}}{G}}\frac{\epsilon}{{\Delta}(\epsilon)} \right)^2 \frac{8\pi G}{3}\frac{\rho}{{\Delta}(\epsilon)} },\nonumber
\end{eqnarray}
now we focus on the asymptotic late time behavior, in this limit $\rho\to 0$, and Eq. \eqref{sol} reduces to
\begin{equation}
H^2+\frac{\kappa}{a^2}\approx \frac{72\pi}{G{\Delta(\epsilon)}^2}\epsilon^2,
\end{equation}
comparing with the Friedmann equation for the de Sitter cosmological model can define an effective cosmological constant $ \Lambda_{eff}$, which to leading order in $\epsilon$ is given by\\
\begin{equation}
 \Lambda_{eff}=\frac{216\pi}{G\Delta^{2}(\epsilon)}\epsilon^{2}.\label{Lambda}           
\end{equation}
This is a surprising result, as we have obtained an effective cosmological constant from the modification to the entropy-area relationship, furthermore, we can trace the origin of this parameter to the SUSY modification to the WDW equation. {Using the current observed value for the cosmological constant 
$\Lambda\sim10^{-52}m^{-2}$, we  fix the value of the parameter $\epsilon\sim 10^{-60}$.} 

From existence of the effective cosmological constant $\Lambda_{eff}$, we can argue entropic or emergent gravity gives plausible explanation to the dark energy problem. 

\section{Final Remarks}\label{conclusiones}
The existence of a  {\it ``dark side of the universe"} is one of the most intriguing problems in physics and its solution  will certainly shake the foundations of physics. Is it a product of our ignorance of the matter content of the universe or is it related to a lack of understanding of the gravitational interaction? In this paper we adopt the former approach. We consider that  dark energy has its origin in modifications of the gravitational interaction,  an that alternative formulation of gravity is warranted. In particular, by attaching an entropic origin for gravity one can study the modifications to the gravitational interaction by studying modifications to the entropy-area relationship. When considering a modified WDW equation derived by using the techniques of SUSY quantum cosmology, we obtained a modified area-entropy relationship.  

In this letter we set our interest on the  origin of dark energy. Using the new entropy-area relationship, we derive a modified Friedmann equation. Focusing our interest in the late-time limit, we find out that the asymptotic behavior is that of a de Sitter cosmology. Then one can define an effective cosmological constant $\Lambda_{eff}$ in terms of the model parameter $\epsilon$. This gives a simple solution to the origin of $\Lambda$ in the context of entropic gravity.

We will like to point out, that in \cite{isaac} using the same entropy-area relationship a modified gravitational force was derived and  used to obtain the correct galactic rotation curves.

{Finally we will like to point out that the volumetric term of the entropy-area relationship that gives the effective cosmological constant $\Lambda_{eff}$ is the same term that gives the correction to the Newtonian gravitational force that accounts for the anomalous galactic rotation curves. Furthermore, the value for $\epsilon$ obtained from Eq.\eqref{Lambda} is consistent with the value needed to explain the anomalous rotation curves of typical galaxies.} This is an uncanny coincidence, that two phenomena at completely different scales can be explained from the same term. The possibility of a deeper connection  between these two aspects of dark sector of the Universe is under study and will be reported elsewhere.

\section*{Acknowledgements}
This work is supported by CONACYT grants 257919, 258982. M. S. is supported by CIIC 28/2018 and by the CONACyT program  ``Estancias sab\'aticas en el extranjero'', grant 31065. J. C. L-D. is supported by UAZ-2016-37235 and UAZ-2015-36952 grants. I. D. S. thanks CONACyT support.
\bibliographystyle{unsrt}
\bibliography{ref}

\end{document}